\documentclass[11pt,a4paper]{article}
\usepackage{epsf,psfig,cite}
\usepackage[latin1]{inputenc}
\begin{document}

\newcommand{\fix}{{\bf FIX}}
\def\<{\langle}
\def\>{\rangle}
\def\={\equiv}
\def\bc{\begin{center}}
\def\bd{b^{\dagger}}
\def\be{\begin{equation}}
\def\bea{\begin{eqnarray}}
\def\bml{\begin{mathletters}}
\def\bt{\begin{tabular}}
\def\e{\epsilon}
\def\ec{\end{center}}
\def\ee{\end{equation}}
\def\eea{\end{eqnarray}}
\def\eml{\end{mathletters}}
\def\et{\end{tabular}}
\def\ff{\frac{1}{4}}
\def\hf{\frac{1}{2}}
\def\hl{\hline}
\def\ni{\noindent}
\def\nn{\nonumber}
\def\np{\newpage}
\def\pr{\partial}

\newcommand{\alphabeta}{$\alpha$-$\beta$\,}
\newcommand{\Prob}[1] {\mbox{P} {\textstyle \left ( #1 \right )}}
\newcommand{\SetOf}[1] {\left\{ #1 \right\}}
\newcommand{\NN}{{\em NN}\,}
\newcommand{\ij}{\< i,j \>}
\newcommand{\mt}[1]{\mbox{\tiny #1}}
\newcommand{\omi}{\omega_i}
\newcommand{\om}{\omega}

\newcommand{\ie}{{\em i.e. }}
\newcommand{\al}{{\em et al. }}

\newcommand{\pfc}{{$\mbox{PF}\overline{\mbox{C}}$}}
\newcommand{\pfs}{{$\mbox{PF}\overline{\mbox{S}}$}}

\newcommand{\Par} { \par \vskip\baselineskip \noindent }

\begin{center}
{\large
{\bf
Automated assignment of SCOP and CATH \\
protein structure classifications from FSSP scores.
}}

\vspace{15truemm}

Gad Getz$^1$, 
Michele Vendruscolo$^2$,
David Sachs$^3$
and Eytan Domany$^1$ \\

\vspace{10truemm}
$^1$ Department of Physics of Complex Systems, \\
Weizmann Institute of Science, Rehovot 76100, Israel  \\
$^2$ Oxford Centre for Molecular Sciences, \\
New Chemistry Laboratory \\
South Parks Road, OX1 3QT Oxford, UK \\
$^3$ Department of Physics, \\
Princeton University \\
Princeton NJ 08544-1117, USA 
\vspace{15truemm}

\end{center}

\date{\today}

\centerline{\large{\bf Abstract}}
\vspace{10truemm}
We present an automated procedure to assign CATH and SCOP classifications to
proteins whose FSSP score is available.
CATH classification is assigned down to the topology level
and SCOP classification to the fold level.
As the FSSP database is updated weekly, this method
makes it possible to update also CATH and SCOP with the same frequency.
Our predictions have a nearly perfect success rate
when ambiguous cases are discarded.
These ambiguous cases are intrinsic in
any protein structure classification,
which relies on structural information alone.
Hence, we
introduce the notion of ``twilight zone for structure classification''.
We further suggest that in order to resolve these ambiguous cases
other criteria of classification,
based also on information about sequence and function,
must be used.

\vspace{25truemm}

\newpage

\section*{INTRODUCTION}

The first step to analyze the vast amount of information 
provided by genome sequencing projects is to organize proteins
(the gene products) into classes with similar properties.
Since during evolution protein structures are much more conserved than
sequences and functions \cite{holm96}, 
proteins are usually classified first by their 
structural similarity (phenetic classification) and then
by the similarity of their sequences 
or by the similarity of their functions (phylogenetic classification)
\cite{thornton99}.

A reliable structural classification scheme is useful for several reasons.
Perhaps the most exciting perspective is the possibility to 
routinely assign a function to newly identified genes \cite{sali98}.
This goal may be achievable as a classified database
provides a library of representative structures to perform
prediction of protein structure by homology
\cite{marti00,heger00} or by threading \cite{bowie91,jones92,fisher96}
and it allows for the identification of 
distant evolutionary relationships \cite{gerstein97}.
In addition, given a particular protein, 
it provides a tool to identify other proteins 
of similar structure and function \cite{murzin96}.
The knowledge of the structure helps to reveal the mechanism
of molecular recognition, involved in catalysis, signalling
and binding \cite{thornton99} and may lead to the rational design of new drugs
\cite{blundell00}.
At a more abstract level, the physical principles dictating 
structural stability of proteins are revealed by their folded state.
Therefore most of the recently proposed methods
to derive energy functions to perform protein fold predictions
rely in different ways on structural data 
\cite{finkelstein97,simons99}.

The most comprehensive repository 
of three dimensional structures of proteins is the 
Protein Data Bank (PDB) \cite{bernstein77}.
The number of released structures is increasing at the pace of about
50 per week and more than 12,000 complete sets of coordinates were 
available at the time of writing.
Many research groups maintain web-accessible hierarchical classifications
of PDB entries.
The most widely used are:
FSSP  \cite{holm97},
CATH  \cite{orengo97},
SCOP  \cite{loconte00},
HOMSTRAD \cite{mizuguchi98},
MMDB \cite{gibrat96} and 
3Dee \cite{siddiqui95}
(See Table \ref{tab:abbrev} for a list of abbreviations).
Here we consider three of these; 
the FSSP, the CATH and the SCOP databases.
Each group has its own way to compare and classify proteins; 
these three classification schemes are, however, 
consistent with each other to a large extent \cite{getz98,hadley99}. 

\subsection*{The FSSP database}

The FSSP (Fold classification based 
on Structure-Structure alignment of Proteins)
uses a fully automated structure comparison algorithm, 
DALI (Distances ALIgnment algorithm) \cite{holm94,dietmann01},
to calculate a pairwise structural similarity measure (the S-score) 
between protein chains. 

The algorithm searches for that amino acid alignment 
between the two protein chains, which yields
the most similar pair of C$_\alpha$ distance maps.
In general, the more geometrically similar are two chain structures, 
the higher is their S-score. The mean and standard deviations of the
S-scores obtained for all the pairs of proteins are evaluated. Shifting
the S-scores by their mean and rescaling by the standard deviation yield
the statistically meaningful Z-scores.   

For classification of structures the FSSP uses 
the Z-scores for all pairs in a 
representative sub-set of the PDB. 
A fold tree is generated by applying an average-linkage 
hierarchical clustering algorithm \cite{jain88}
to this all-against-all Z-score matrix.
An alternate classification based on a more common
four-level hierarchy is also available \cite{dietmann01}.

\subsection*{CATH database}

Thornton and coworkers use a combination 
of automatic and manual procedures to create a
hierarchical classification of domains (CATH) \cite{orengo97}.
They arrange domains in a four level hierarchy of families 
according to the protein class 
({\em C}\ ), architecture ({\em A}\ ), topology ({\em T}\ ) 
and homologous superfamily ({\em H}\ ).
The class level describes the secondary structures found in 
the domain \cite{levitt76} and is created {\em automatically}. 
There are four class types: 
mainly-$\alpha$, mainly-$\beta$, \alphabeta and proteins with few
secondary structures (FSS). 
The architecture level, on the other hand, is assigned {\em manually} 
(using human judgement) and describes the shape 
created by the relative orientation of the secondary structure units. 
The shape families are chosen according to a commonly 
used structure classification,
like barrel, sandwich, roll, etc. 
The topology level groups together all structures with 
similar sequential connectivity between their
secondary structure elements.
Structures with high structural and functional similarity are put in the same 
fourth level family, called homologous superfamily.
Both the topology and homologous superfamily levels 
are assigned by thresholding a 
calculated structural similarity measure (SSAP) 
at two different levels, respectively 
\cite{taylor89,orengo92}.
The CATH database has been recently
linked to the DHS (Dictionary of Homologous Superfamilies) database 
\cite{bray00} which allows to further analyze structural 
and functional features of evolutionary related proteins.
There is a growing need for annotating proteins classified in structural
databases as structural genomic initiatives are providing a large number
of new proteins whose function might be gathered by distant homology
informations.

\subsection*{The SCOP database}
The SCOP (Structural Classification of Proteins) \cite{loconte00} database
is organized hierarchically. The lower two levels (family and
superfamily) describe near and distant evolutionary relationship,
the third (fold) describes structural similarity and the top level (class)
the secondary structure content \cite{levitt76}.
SCOP is linked to the ASTRAL compendium \cite{brenner00}, which provides
a series of tools for further analysis of the classified structures,
mainly through the use of their sequence.
At variance with FSSP and CATH, SCOP is constructed manually, by visual
inspection and comparison of not only structures 
but also sequences and functions. 

\subsection*{Automated assignment of SCOP and CATH classifications}
In this work we present a method, {\em Classification by Optimization} (CO),
to predict without human intervention
the SCOP fold level and the CATH topology level from the
FSSP pairwise structure similarity score.
A protein for which the Z score is available is classified into
a SCOP fold and into a CATH topology by the CO method,
an optimization procedure which finds the assignment of minimal cost, 
where the cost is defined in terms of Z scores (see Methods).
The query for the classification of any such protein can be submitted
to the web site \cite{getz01web}.

\section*{RESULTS}

\subsection*{Consistency of the FSSP, CATH and SCOP classifications}

We found that the FSSP and CATH databases are consistent \cite{getz98}.
In this section we show that SCOP is also consistent with these to a large
extent (see also \cite{hadley99}).
In the rest of this work we will use this fact to derive an automated procedure
to assign the CATH and SCOP classifications starting from the FSSP Z scores
(which are updated weekly) in a fully automated fashion to include new releases
in the PDB \cite{holm96}.
Here we further discuss 
the consistency of the three classification schemes 
by introducing concepts and quantities that will be later used
in the prediction of the CATH and SCOP classifications.

\begin{figure}
    \centerline{ 
       \psfig{figure=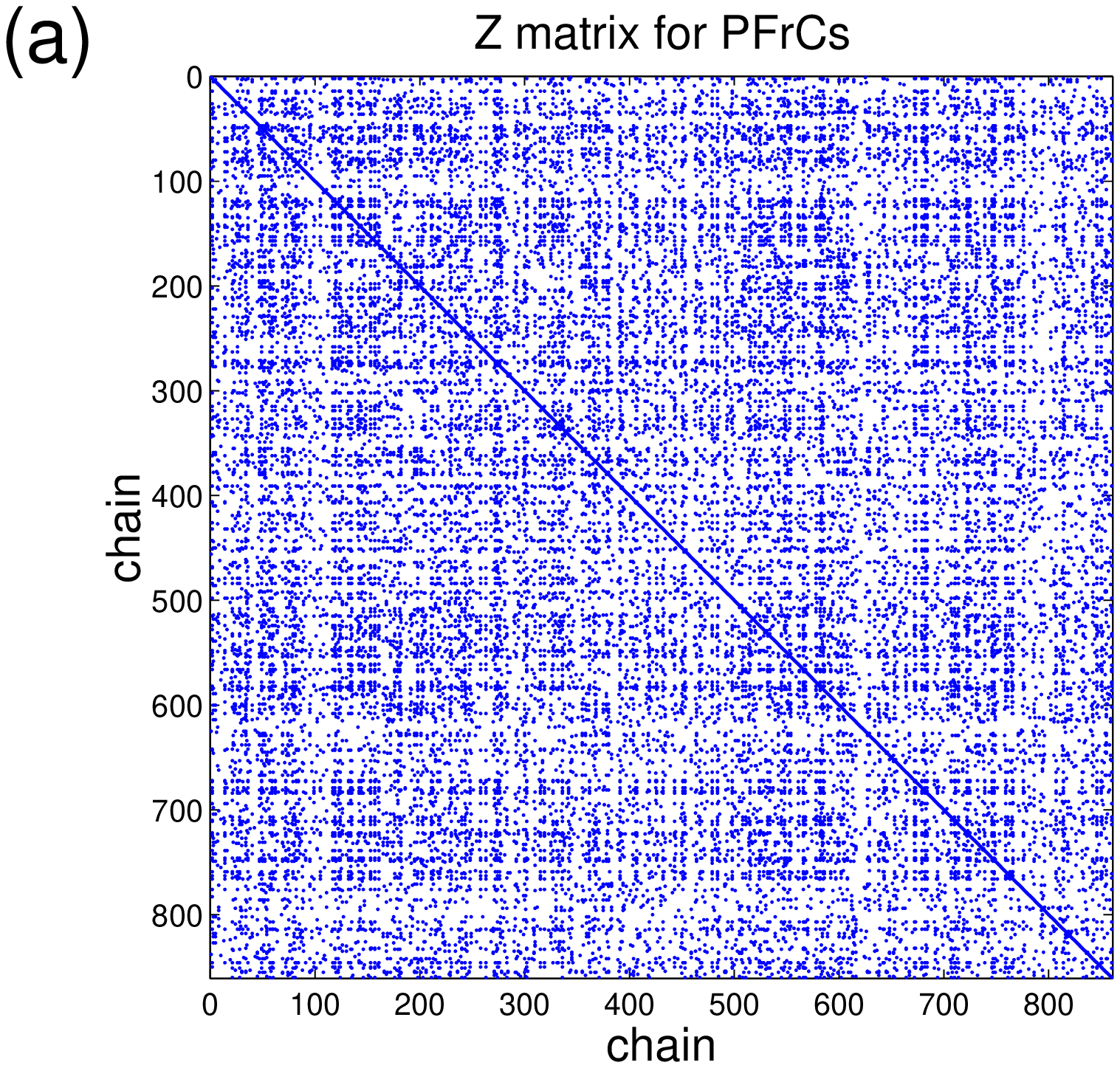,width=8.0cm} 
       \hspace{1.0cm} 
   	\psfig{figure=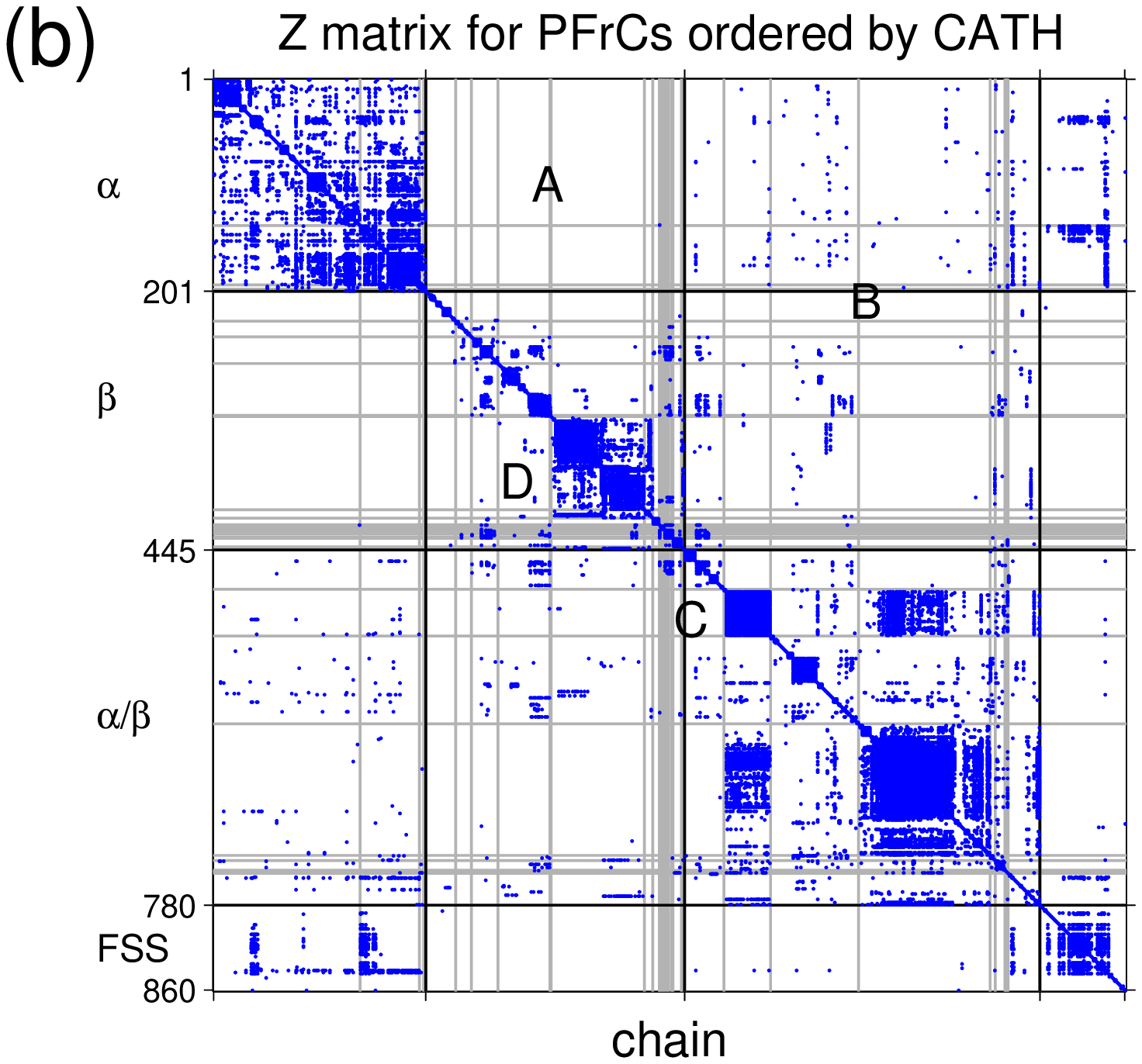,width=8.0cm} 
     }
\caption{(a) Z-score matrix between all pairs of proteins in the PFrCs set. 
A black dot represents $Z>2.0$.
(b) Same Z-score matrix with rows and columns rearranged by using the CATH
classification (see text).
Part (b) shows the underlying order behind the apparent randomness 
of part (a) and illustrates the extent to which
the FSSP Z-scores reflect the CATH classification.
The regions A, B, C and D are discussed in the text.
}
\label{fig:zcath}
\end{figure}

We first illustrate the correlation between
the FSSP similarity score and the CATH classification.
A simple and visually appealing way to study this problem is shown
in Figure \ref{fig:zcath}.
The element $Z_{ij}$ of the Z-score matrix (Figure \ref{fig:zcath}a) 
represents the score for superimposing structure $i$ with structure $j$
of the set PFrCs (a subset of the proteins in FSSP and CATH, see 
Table \ref{tab:prot_sets} and Methods) 
using the DALI algorithm \cite{holm94,dietmann01}. 
In Figure \ref{fig:zcath}a only the pairs with $Z > 2$ are shown,
therefore the matrix is sparse and the proteins are ordered 
in a random fashion.
Figure \ref{fig:zcath}b is produced by reordering the rows and 
columns of the original Z-score matrix (Figure \ref{fig:zcath}a).
The reordering is performed according to the CATH classification
in the following way:
For each of the proteins in this set we have the CATH
classifications at all levels.
First, we order the proteins by their class; 
within the class, by the architecture; 
within it by the topology and so on. 
This reordering generates a permutation of the
columns and rows of the Z matrix.
The solid black grid in Figure \ref{fig:zcath}b separates
the proteins according to their CATH class and a thin grid is
placed at the boundaries between architectures. 

Figure \ref{fig:zcath}b shows the underlying order
behind the apparent randomness of Figure \ref{fig:zcath}a
and reveals the extent to which
the FSSP Z-scores reflect the CATH classification.

Several interesting observations can be made.
First, consider the Class level of CATH.
As can be seen in Figure \ref{fig:zcath}b,
there are no matrix elements with $Z>2.0$ in region A,
that connects proteins of the
mainly-$\alpha$ class to the mainly-$\beta$ class.
At variance with this,
some proteins from both of these classes have large Z scores
with proteins from the $\alpha$-$\beta$ class (region B).
This is reasonable, because of the way similarity is defined by
FSSP; a mainly-$\alpha$ protein can have a high Z-score with
an $\alpha$-$\beta$ protein, due to high similarity with the
$\alpha$ part.
Second, turning to the Architecture level, we observe
that there are architecture families which are highly connected within
themselves, e.g. $\alpha$-$\beta$ barrels (482-525: region C),
whereas for others the intra-family connections  are more sparse.
The similarities within
the mainly-$\beta$ sandwich family (318-406: region D) have two
relatively distinct subgroups
which suggest an inner structure corresponding to the lower levels
in the CATH hierarchy.
Checking the topology level (the third CATH level)
for this architecture, one indeed finds two large topology
sub-families, the immunoglobulin-like proteins
(324-366: upper left part of region D) and the
Jelly-Rolls (373-402: lower right part of region D),
which correspond precisely to the two strongly connected sub-groups that
appear in Figure \ref{fig:zcath}b.

We found that
the CATH classification at the level of topology is reflected
in the Z-matrix. This is to be expected since the Z-score measures
the structural similarity of two aligned proteins, while preserving
their connectivity.
Overall, this analysis shows that the Z-matrix is correlated with the
CATH classification.
In a similar way it is possible to show that the Z score
is correlated with the SCOP classification.
The results are available at the web site \cite{getz01web}.

These findings suggest that $Z$ scores can be used to 
predict the CATH and SCOP classifications of yet unclassified proteins.
In what follows, we will demonstrate that this indeed can be done. 
We also estimate the success rate of our predictions and provide a web 
site \cite{getz01web} that can be used to retrieve our predictions
for the CATH topology and the SCOP fold for new entries in FSSP.

\begin{figure}
    \centerline{ \psfig{figure=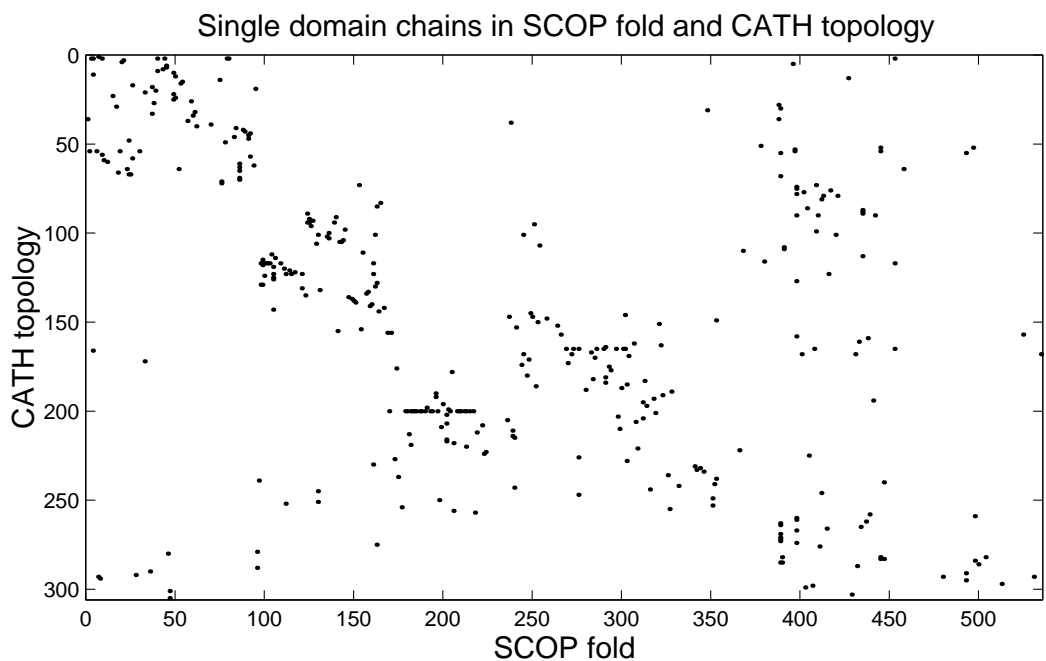,width=14.0cm} }
\caption{Translation table from the CATH topology to the SCOP fold and {\em vice versa}.
Non-zero entries of $\widehat{T}_{ij}$ appear as black dots. $\widehat{T}_{ij}$ is proportional to the number of proteins of CATH topology $i$ that have a SCOP fold $j$ in PCsSs.
}
\label{fig:translation_table}
\end{figure}

We also verified that the CATH and SCOP classifications are to a large
extent mutually compatible.  An immediate consequence of this is that it 
is possible to construct a ``translation table'', $\widehat{T}$, from the proteins
that have already both a CATH and a SCOP classification. 
In this way,
given a CATH entry one can obtain the corresponding SCOP classification
(see Fig. \ref{fig:translation_table}).
Row $i$ of the table refers to a particular CATH topology
and column $j$ to a particular SCOP fold.
The element $\widehat{T}_{ij}$ of the table is the measured 
fraction of times that a protein has a CATH topology $i$ and a SCOP fold $j$. 
This number is calculated by enumerating all the 10197 single-domain proteins
with known CATH and SCOP classifications (PCsSs) and it is an estimate of 
$T_{ij}$, the joint probability distribution for a protein to have 
CATH topology $i$ and SCOP fold $j$.
Had the CATH and SCOP classifications been independent,
every element $T_{ij}$ could have been expressed as a product
of $C_i$, the fraction of proteins that belong to CATH topology $i$,  and 
$S_j$, the fraction that belongs to SCOP fold $j$, \ie $T_{ij}=C_i*S_j$.
Randomly placing 10197 proteins using such a probability distribution
yields $4780 \pm 40$ non-zero elements in the matrix.
In the other extreme case, had there been a full correspondence between 
the SCOP and CATH 
classifications, the table would have had a single non-zero element in each 
row and column 
(in each CATH topology row the non-zero element 
would have been in that SCOP fold column which
corresponds to it). In this case, the proteins in PCsSs would have been 
distributed among 284 non-zero elements (the number of distinct CATH 
topologies in PCsSs).

We found 369 
non-zero elements in $\widehat{T}$, meaning that 
the CATH and SCOP classifications are highly dependent. 
Still, the correspondence is not 
entirely one-to-one;
in general, more than one SCOP fold corresponds to a given CATH topology.
The number of such folds is, however, typically small.
Such a translation table may be used to predict the SCOP
classification of a structure already classified in CATH,
or at least to significantly restrict the number of possibilities, 
(and {\it vice versa}).
For example, the assignment of the CATH topology to a 
protein with known SCOP fold can be done by selecting the CATH 
topology with the largest value in the translation 
table for that particular SCOP fold. Such an assignment is correct in 93\% of 
the cases. The corresponding assignment of the 
SCOP fold from the CATH topology is correct in 82\% of the cases.
Although this is possibly useful information,
in this work we do not assign classifications in this way.

\section*{Summary of the CO Classification Performance}
Every time the FSSP Z scores are updated (once a week) the CO classification
can be applied to all the proteins that appear in the new FSSP release
but are not yet classified in CATH or in SCOP.
The possible outcomes of the classification procedure are:

\begin{enumerate}
\item
{\it Correct classification:} 
the predicted classification will agree with
the future release of the databases. 
\item
{\it Rejection:} the program is unable to classify the structure.
\item
{\em Ambiguous classification:} 
a classification is returned (both for CATH and SCOP)
but a later release provides a different classification.
\end{enumerate}

The frequencies of these outcomes greatly depends on the statistics
of the set of proteins to be classified.
More specifically, rejected proteins are of two types: 
proteins that do not have high Z score with any other proteins
(``islands'' - see Methods)
and clusters of proteins that are similar among themselves but do not have
high Z score with other proteins outside their cluster (``super-islands'').
The fraction of islands and super-islands is a feature of the 
particular set of proteins to be classified. 
The occurrence of a super-island suggests that a new classification type 
(a new topology for CATH and and new fold for SCOP) might be needed.
The work of maintaining CATH and SCOP can be thus focused on 
the classification of a representative from each of these super-islands.

For the set PFCs the fraction of islands and superislands is 5\%. 
We used this set
to provide an upper bound for the performance of the CO method (see below);
however, for the set \pfc~ the fraction of rejections goes up to 22\%.
If rejections are not counted, we classify correctly 98\% of the PFCs proteins.
On the other hand, we could test our predictions also against 
the new CATH release v2.0. Out of  1582 proteins that were assigned to
previously existing CATH topologies, 
CO has classified correctly 80\%. 
The difference in success rates between PFCs and \pfc~ is due to the different
way in which the test set is nested in the larger set of structures with known 
classification. In the first case the test set consisted of
20\% of the members of PFCs, selected at random; the remaining 80\% were used 
to ``predict" the classification of the test set. 
In the second case, the members of CATH v1.7 were used 
to predict the classification of the new proteins that were 
added when CATH v2.0 was released. 
These new structures are {\it not} distributed uniformly at random
among the members of CATH v1.7.

Ambiguous classifications are due to two different mechanisms. The first stems
from a well known problem with the way the FSSP similarity index is calculated
(the ``Russian doll effect" - see below). The second kind of ``mistake" is 
actually {\it not} a wrong classification; rather, it happens when the newly
classified structure lies within the ambiguous ``twilight zone" between two 
closely related topologies (for CATH) or folds (for SCOP), as demonstrated in
detail below.

\subsection*{Automated assignment of CATH classification from FSSP}

In this section we describe the procedure that we used to predict CATH topology 
level from the FSSP scores. 
We identified a set of 7431 proteins (\pfc~, see Methods) that appear
in FSSP but were not yet processed by CATH 1.7.
Our goal is to predict the CATH topology of these 7431 proteins by using 
(1) the Z-scores between all proteins in PF (see Methods) and 
(2) the known classifications of the set PFrCs (see Methods).

Predicting topologies is a classification problem
which we treated with pattern recognition tools.
We tested several prediction algorithms using
cross-validation to estimate their performance \cite{getz98}. 
Every one of the algorithms that were tested
can be viewed as a  two-stage process.
In the first stage a new similarity measure is produced from
the original Z-scores. This is done either by a direct rescaling of the
original Z-scores, or by using the results of various hierarchical clustering
methods to produce new similarity measures.
The second stage consists of using these similarities
as the input to some classification
method, yielding predictions for the classes and architectures.
In this work we present only results obtained by one particular method - CO - 
which uses the 
original Z score as a similarity measure
(see Methods). A complete list of the results obtained by using
other methods can be found in \cite{getz98} which is available on the 
web site \cite{getz01web}.

\begin{figure}
    \centerline{ 
      \psfig{figure=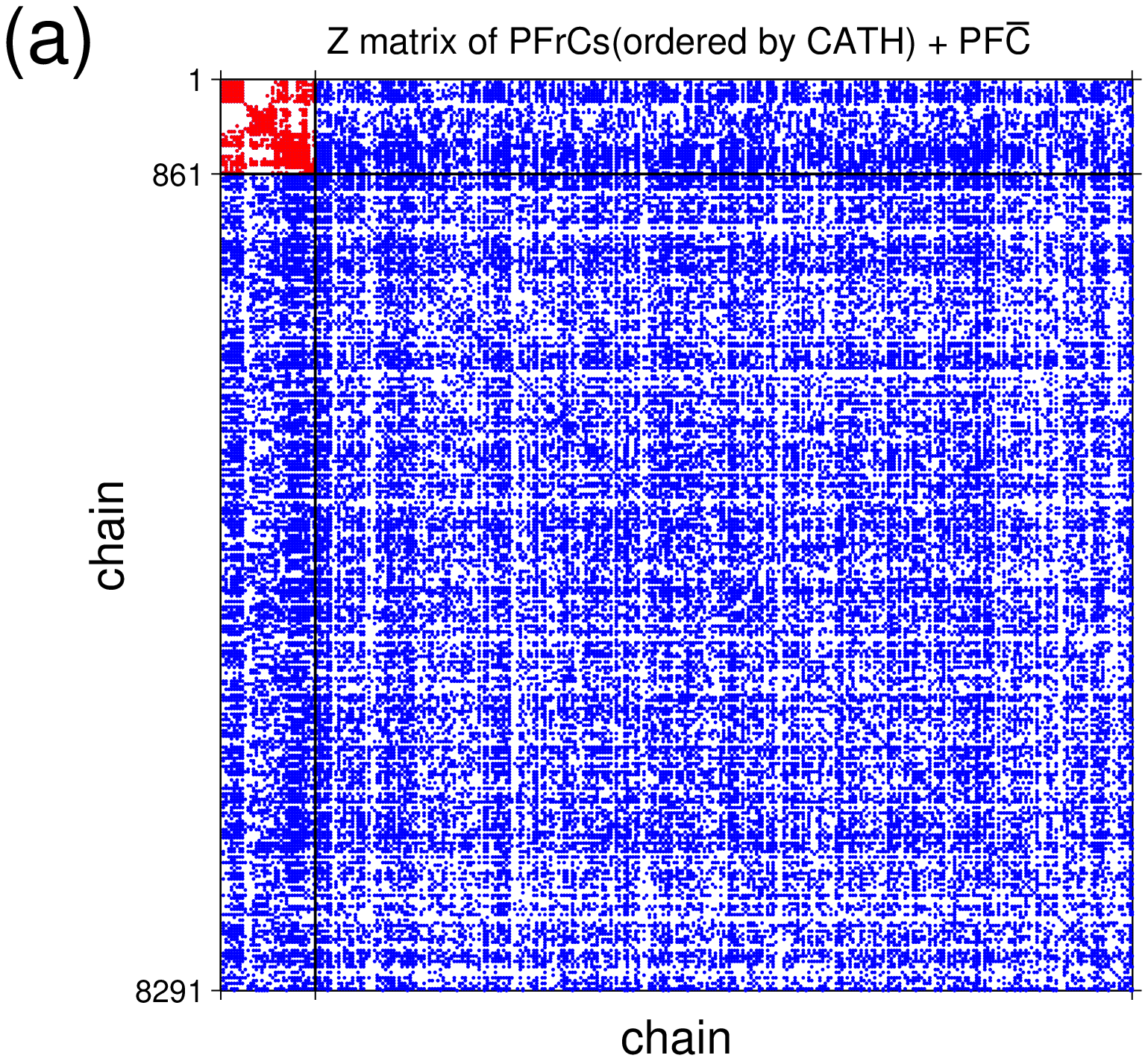,width=8.0cm}
       \hspace{1.0cm}
       \psfig{figure=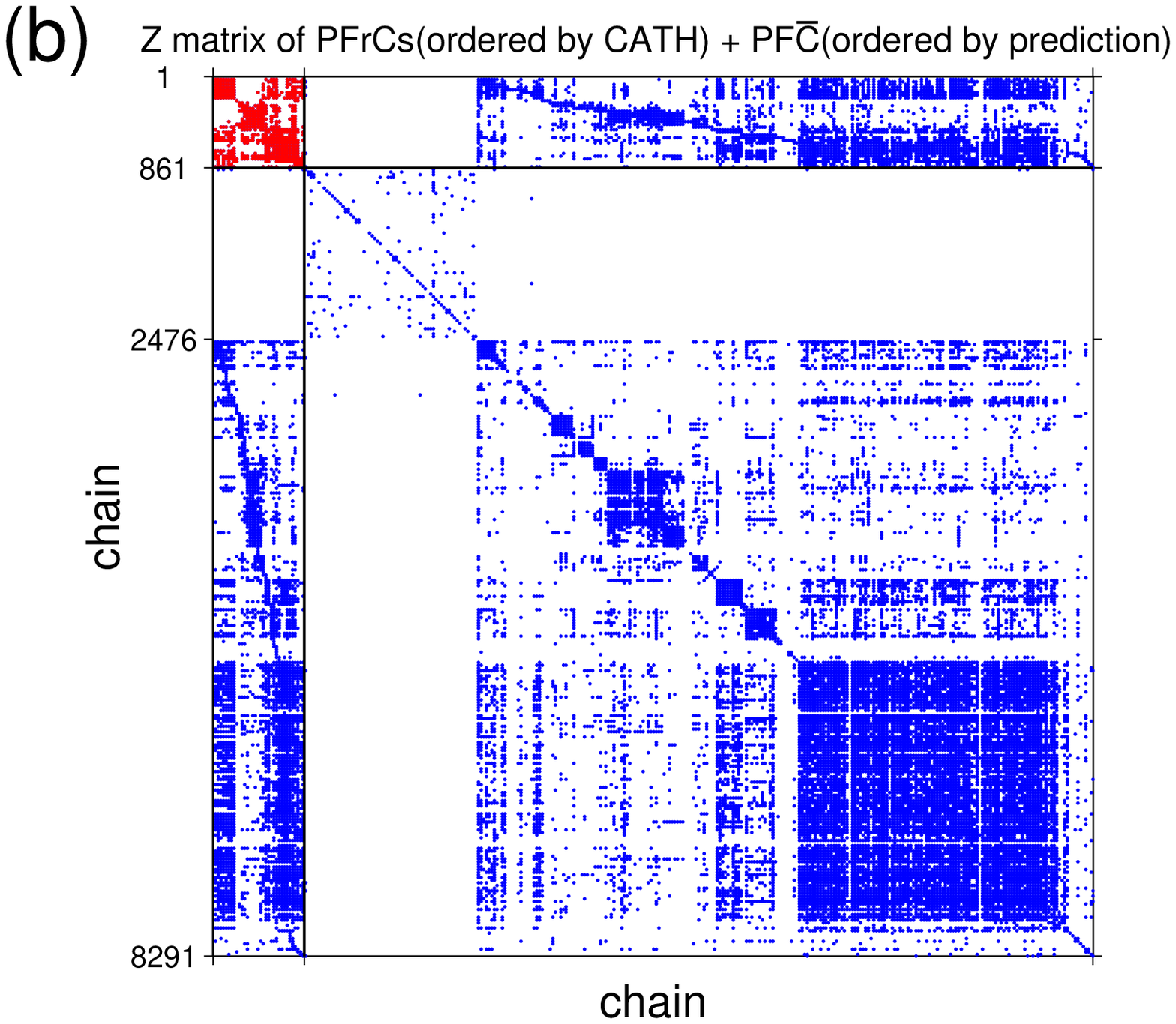,width=8.45cm}
	}
\caption{(a) Z-score matrix between all pairs of proteins in the combined 
PFrCs+\pfc~ sets. 
The submatrix in the upper left corner is the
reordered Z score matrix of the set PFrCs which was already shown 
in Figure \protect \ref{fig:zcath}b.
The rest of the matrix presents the Z scores for the proteins in the set \pfc~. 
(b)
The same matrix as in (a) with
the rows and columns relative to the proteins in \pfc~
reordered according to our assignment of their CATH topology.
With the CO method the original order in the submatrix PFrCs 
is propagated to the entire matrix.
}
\label{fig:zcath_pred}
\end{figure}

Our final assignments for the set \pfc~~ using the CO method are 
listed in the web site. 
A more illustrative way to present these results is shown in 
Figure \ref{fig:zcath_pred}. 
In Figure \ref{fig:zcath_pred}a we present the Z score matrix for the combined
set PFrCs+\pfc~.  The submatrix in the upper left corner is the
reordered Z score matrix of the set PFrCs which was already shown 
in Figure \ref{fig:zcath}b.
The rest of the matrix in Figure \ref{fig:zcath_pred}a presents the Z scores
of PFrCs with the set \pfc~ 
(randomly ordered) and the Z scores of \pfc~
among themselves.
In Figure \ref{fig:zcath_pred}b
we reordered the rows and columns whose index $> 860$, corresponding to 
proteins in \pfc~. Whereas 
in the matrix of Figure \ref{fig:zcath_pred}a these proteins appear in a random
order, in \ref{fig:zcath_pred}b they appear in the order imposed by 
our prediction of their CATH topology.
One can see that the original order in the submatrix PFrCs
is propagated by our assignment procedure to the set \pfc~.
For example, focus on the small black square at the upper left corner of
the matrix. This small black square represents the high Z-scores among the
mainly-$\alpha$ class of proteins in
PFrCs. In the corresponding top rows of the full matrix we see high Z-scores 
between these structures and some proteins from \pfc~. In particular, the small
group with indices near 2476, are ``close" to these mainly-$\alpha$ structures,
and hence are also classified as such. On the other hand there is a large group
of structures from \pfc~ (between 861 and 2476), 
which do not have high Z-scores
with {\it any} of the proteins in PFrCs or with any of the other structures in
\pfc~ with index $> 2476$. Hence we are unable to classify this group of
structures on the basis of their FSSP scores. 

Figure \ref{fig:zcath_pred}b illustrates the central idea of this work.
We perform a task which is intermediate between clustering and classification.
We take proteins of known classification and we use them as fixed 
{\em a priori} values in a clustering procedure.
 
The overall success rate of our prediction estimated by 
cross-validation was 93\%. 
In order to understand the significance of these success rates we
derived a statistical (see Methods) upper bound for this kind of prediction.
This upper bound is 95\% (see Methods), hence the figure of 93/95=98\% 
given above.\footnote{ 
One must keep in mind that the estimated success rate is calculated 
for all proteins; both FSSP representatives 
($\approx 10\%$ of the proteins) and non-representatives.
Since the presence of homologous proteins can create a bias in these estimates,
we also tested the success rate of predicting the CATH topology only for the 
FSSP representatives which yielded 63\%, to 
be compared with the corresponding upper bound of 74\%.}

We estimated the accuracy of the prediction by using the following procedure.
First the set PFCs was randomly ``diluted''; that is, we randomly chose a
certain fraction of the proteins in PFCs 
and placed them in a test set, pretending
that we do not know their classification.  The FSSP scores of the entire
set were then used to classify the test set. For each protein from the test
set we either return a predicted classification or reject the protein (i.e.
we declare that we are unable to classify it). 
The quality of any classification algorithm (see Methods) is measured by its
success rate (fraction of correctly classified proteins, out of the
test set) and by the purity  (success rate out of the non-rejected
proteins). 
For the CO method the results were 93\% for the success rate and 98\% for
the purity (using a dilution of 20\%).
More extensive tests at other dilutions and for other methods are
of classification are discussed in \cite{getz98} 
and available at the web site \cite{getz01web}.

We also tested directly the reliability of the CO assignments by using the CATH
version 2.0 (PC2). In PC2, 1640 single domain proteins that are present 
in \pfc~ were assigned to one of the topologies that existed in v1.7. 
58 of these we ``rejected''.
In 1266 cases out of the remaining 1582 (80\%) our prediction 
agrees with the one given in CATH v2.0.
Almost all the cases in which we misassigned a domain 
can be explained in a simple way.
These cases are discussed in detail in a following section.

The CO method can also be used to predict directly the C level 
and the A level of CATH.
We found that when the C and A levels were predicted as a byproduct of
predicting the T level, the resulting C and A were consistent with
those predicted directly.

\subsection*{Automated assignment of SCOP classification from FSSP}

We used the CO method to predict the SCOP fold for a set of 3451
proteins (\pfs) that belong PF but not yet in PS.
The results are available on the web site \cite{getz01web}.
The estimated success rate (by cross-validation) was 93\%. 
As in the case of CATH this number increased when we discarded 
proteins in the ``twilight zone'' (see next section).

\subsection*{Twilight zone for protein classification}
\begin{figure}
    \centerline{ \psfig{figure=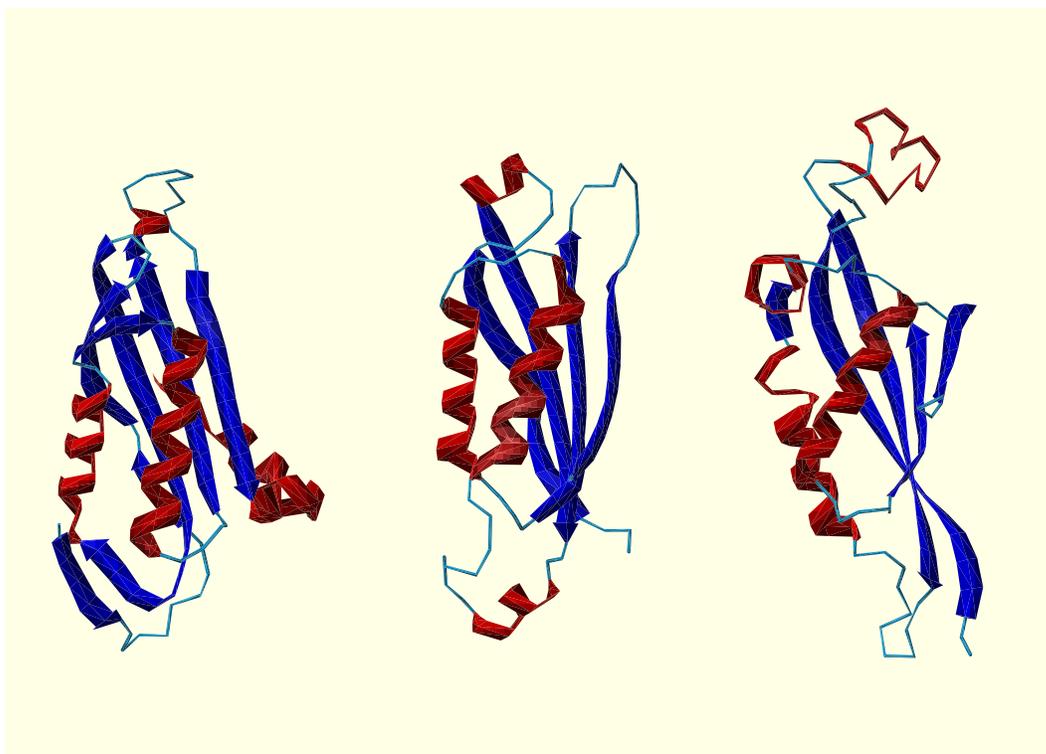,width=14.0cm} }
\caption{(centre) 
Protein 1dhn which has a CATH $\alpha~\beta$ 3-layer ($\beta\beta\alpha$)
sandwich Aspartylglucosaminidase chain B (3.50.11) topology.
(left)
Protein 1a8rA which has also a CATH $\alpha~\beta$ 3-layer ($\beta\beta\alpha$)
sandwich Aspartylglucosaminidase chain B (3.50.11) topology
and has Z score of 9.3 with protein 1dhn.
(right)
Protein 1b66A which has a CATH $\alpha~\beta$ 2-layer
sandwich Tetrahydropterin Synthase, subunit A  (3.30.479) topology
and has Z score of 8.7 with protein 1dhn.
This example illustrates how 
structural information alone might be insufficient to provide
a clear-cut criterion for the classification of this protein.
}
\label{fig:ambiguity}
\end{figure}

\begin{figure}
    \centerline{ \psfig{figure=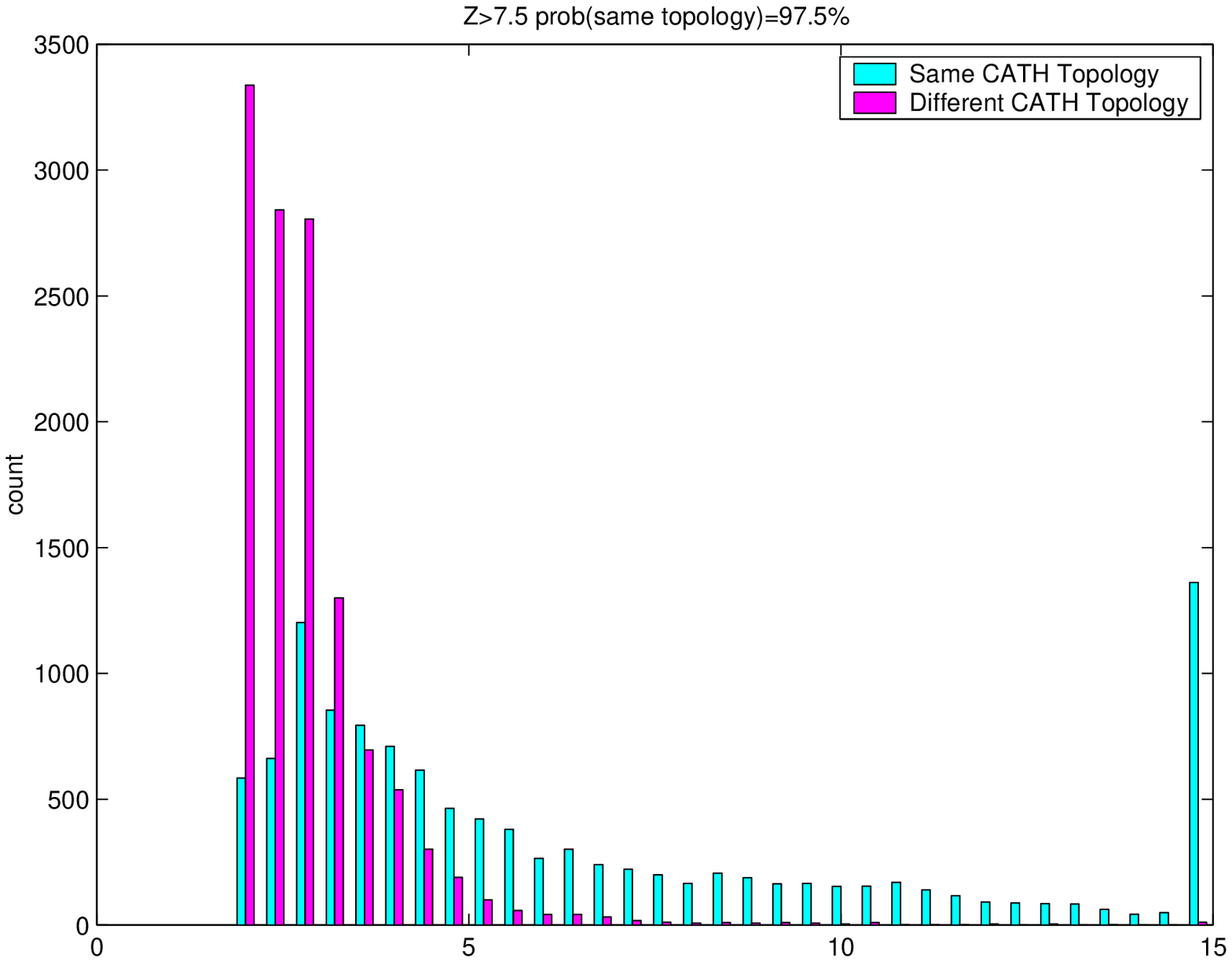,width=8.0cm}
                 \psfig{figure=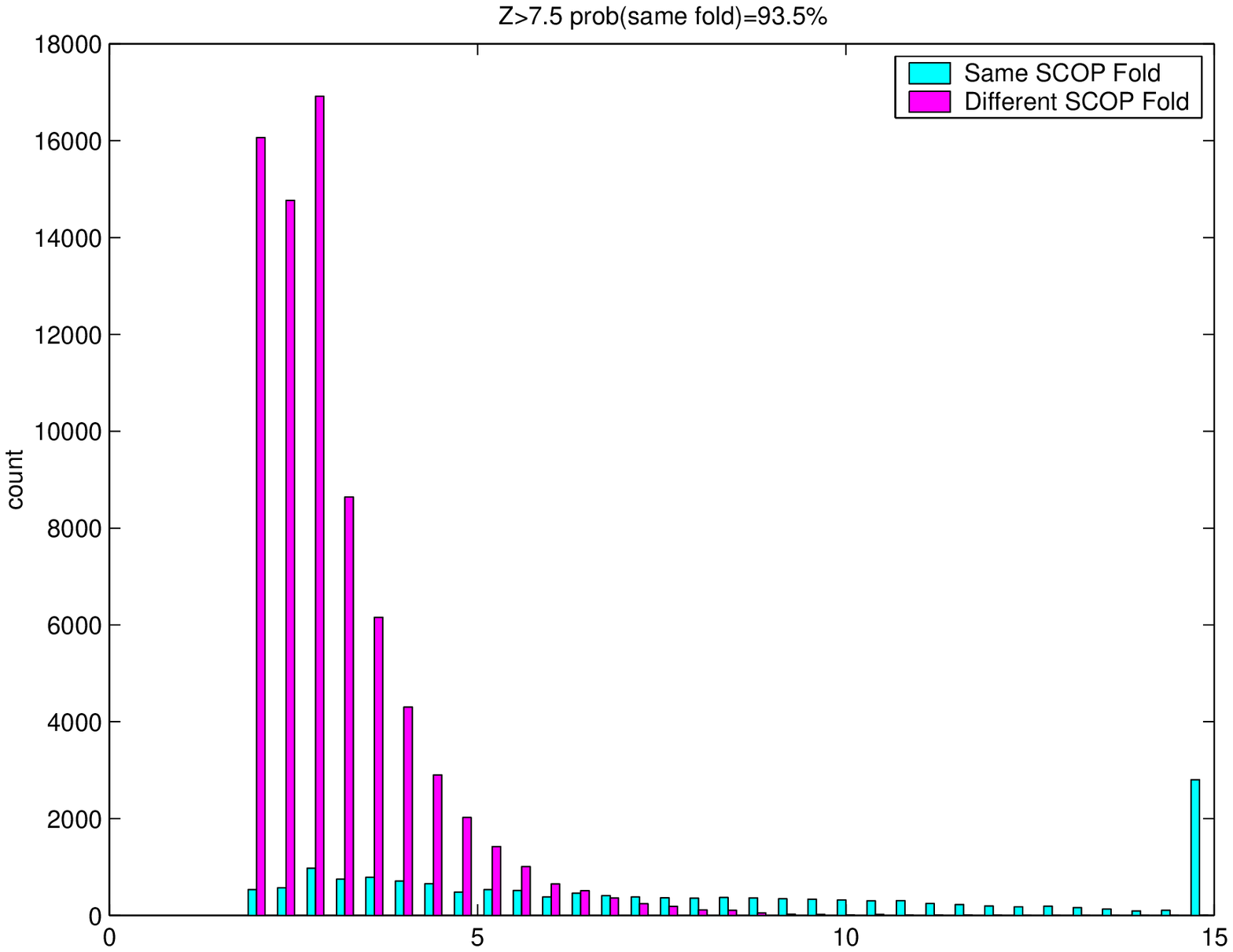,width=8.0cm }}
\caption{Twilight zone for protein structure classification.
(a)
The number of protein pairs of with a given FSSP $Z$ score that have
different CATH folds is a rapidly decaying function of $Z$.
On the contrary, the number of proteins pairs with the same CATH fold
is decaying slowly. For $Z<5$ there is a non-negligible probability
to have different folds. We call this threshold the ``twilight zone
for structure classification''.
(b) The corresponding histogram for SCOP folds. The number of SCOP folds
is larger than the number of CATH topologies, hence the twilight zone is
$Z \simeq 7$.
}
\label{fig:twilight}
\end{figure}

The attempt to assign a new protein to a known fold might lead to frustration
because at times one is undecided about two or more possibilities.
In order to assess that two proteins have similar structures,
a similarity score is needed. FSSP uses the Z score, CATH uses the SSAP score
and SCOP a subjective evaluation, which is also a kind of score.
The problem arises when the protein to be classified has high scores
with two proteins already classified, but to different topologies.
In this paper these proteins are called ``borders'' (see Methods).
Being or not a border protein depends on the similarity score.
We showed, however, that FSSP, CATH and SCOP are to a large extent consistent
classifications.
We therefore suggest that there are ``intrinsically'' ambiguous cases --
cases that are unavoidable in structure comparison.
We refer to these ambiguous regions in structure space as 
the ``twilight zone'' in analogy with the case of
protein sequence comparison where
proteins with sequence similarity below 30\% can not be reliably assigned
to the same fold.
We illustrate this concept by 
a typical case, shown in Fig. \ref{fig:ambiguity}. This is a border protein.
Protein 1dhn (the central one) is the one to be classified
(in fact, it is a 3-layer sandwich according to CATH).
It has a Z score of 9.3 with protein 1a8rA (on the left) 
which is a 3-layer sandwich topology
and a Z score of 8.7 with protein 1b66A (on the right) which is a 
2-layer sandwich topology. This example illustrates how 
structural information alone might not provide
a clear-cut criterion for classification of this protein.
The incidence of the twilight zone is shown in Fig. \ref{fig:twilight}.
In Fig. \ref{fig:twilight}a we present the histogram of the number of
protein pairs that have different CATH topologies as a function of their
$Z$ score. This number is a rapidly decaying function of $Z$.
On the contrary, the number of pairs with the same CATH topology
is a slowly decaying function of $Z$. For $Z>3$ the probability
of having the same CATH topology becomes greater than that of having
different topologies.
For $Z>7.5$ the probability to have the same topology is 97.5 \%.
In \ref{fig:twilight}b we show the corresponding figure for SCOP.
The number of folds in SCOP is larger than the number of topologies
in CATH, therefore there is more ambiguity.
However, also in this case for $Z>7.5$ the probability
to have the same topology is 93.5 \%.
Taken together these results indicate that the twilight zone
for structure comparison can be bound by  $Z \leq 7$.

There are other cases in which the classification of a particular
protein is inconsistent with that of all its neighbors.
For example, proteins that we called ``colonies'' (see Methods)
are such that none of their neighbors are of their own kind.
This means that the FSSP scores imply that these proteins are
similar only to proteins of different classes and architectures.
Identifying these proteins can also focus the attention to possible
misclassification or to drawbacks of the Z-score.
For example, one of the 49 colonies (at the architecture level) 
that we found in CATH is the PDB entry
1rboC, which is classified as a $\alpha$-$\beta$ two-layer sandwich. 
It has 15 neighbors in PC, 14 of which are classified as
mainly-$\beta$ sandwiches. 

We summarize the results about the assignments of the
CATH architecture for proteins that already have a CATH classification (PFCs)
in a ``confusion table'' (see Table \ref{tab:confusion}).
The first column lists the ``correct" classification (as given in CATH v1.7
for the test set); the second column gives the assignments by CO (correct,
incorrect or reject) and the third column lists the corresponding percentages.
A full list of the inconsistent proteins is available on the web 
site \cite{getz01web}.

Another problem is that there are some large Z-scores
between proteins of different architectures.
Such large Z-scores arise when
a protein of one particular architecture
has a similar structure to a part of a protein
of a different architecture.
Swindells \al call the phenomenon of structures within structures,
the ``Russian doll'' effect \cite{swindells98}.
Such cases are common between architectures of long proteins
that contain sub-structures corresponding
to architectures of shorter proteins,
{\em e.g.} there are many two-layer sandwich proteins that resemble a part of
three-layer sandwich proteins.
Such relationships can occur at the class level,
{\em e.g.} \alphabeta proteins that contain
mainly-$\alpha$ or mainly-$\beta$ proteins (1rboC, 1hgeA). They can also
occur at the architecture level within the same class,
{\em e.g.} \alphabeta complex architecture contains
\alphabeta two-layer sandwich (1regX).
Other inconsistencies occur when proteins
fit two architecture definitions. 
\subsection*{Class prediction using the web site}
In order to retrieve our prediction for the CATH topology or SCOP fold 
of a protein, one can use the web site \cite{getz01web} 
by entering the protein chain identifier 
in the search box and submitting the query. 
If the protein appears in our database then  
a table will be returned containing both the known and 
the predicted SCOP and CATH classifications. 
For example, the submission of the chain identifier ``1cuoA'' returns 
Table \ref{tab:1cuoA}. This protein was classified 
by neither CATH v1.7 nor SCOP 1.53 which are the basis of our predictions. 
We predicted it to belong to CATH topology 2.60.40 
and SCOP fold 2.5. Later, the release CATH v2.0 identified 1cuoA as 2.60.40.

\section*{CONCLUSIONS}
The rapidly increasing number of experimentally derived
protein structures requires a continuous updating of the existing
structure classification databases.
Each group adopts different classification criteria
at the level of sequence, of structure and of function similarities.
A comparison between different classification schemes
can help understanding the optimal interplay between different levels,
it can reveal possible misclassification,
and it can ultimately offer a fully automated updating procedure.
Manual steps can be automated in an ever-increasing way
by using the tools made available by other databases.

In this work we showed that it is possible to automatically 
predict the CATH topology and the SCOP fold from the FSSP Z-scores. 
It is possible to submit a protein of 
unknown CATH or SCOP classifications but known FSSP Z-scores to the web site 
\cite{getz01web} to obtain its CATH and SCOP classifications.
Since the FSSP database is updated weekly, our procedure
offers the possibility to update also CATH and SCOP with the
same frequency (at least down to the topology and fold level, respectively).
We introduced a classification method that clusters together 
structures of known and unknown classification
according to their Z-scores.
When proteins outside the twilight zone for structure comparison
are considered our method is highly reliable.
We suggest that, in order to classify proteins within the twilight zone,
other classification criteria, based on sequence and function similarity,
must be adopted.

The advent of genome projects is multiplying the efforts
in the field of protein classification.
In the past the aim was to help finding the structure of the particular protein
which was interesting at a given time.
Now the hope is to find a large representative set of structures
which can encompass most of the existing folds, 
possibly all of them \cite{sali98}.
In such a large scale project human intervention, which is precious
in setting the principles of classification, should be gradually replaced by
automated procedures.

\section*{Acknowledgments}

We thank Liisa Holm for making the raw FSSP data available to us and for
useful discussions during the initial stages of this project. 
This work is based on a thesis for the M.Sc.
degree submitted by G. G. to Tel-Aviv University (1998). We also thank Noam
Shental for discussions.
This research was supported by grants from the Minerva Foundation,
the Germany-Israel Science Foundation (GIF) and
the US - Israel  Science Foundation (BSF).
M. V. is supported by an European Molecular Biology Organization (EMBO)
long term fellowship; he also
thanks the Einstein Center for Theoretical Physics for partial
support of his stay at the Weizmann Institute.
D. S. thanks the Weizmann Institute of Science for hospitality
while part of this work was carried out.

\section*{MATERIALS AND METHODS}

\subsection*{Databases and protein sets}

As the CATH and SCOP databases classify domains, 
whereas FSSP deals with chains,
we considered only chains which form a single domain and,
therefore, these proteins appear as a single entry in the three databases.
Several groups have developed methods to identify protein domains
\cite{holm94,islam95,swindells95,siddiqui95,sowdahamini96}.
In this work, we used the Dali Domain Dictionary \cite{dietmann01}
to identify single domain proteins.

We used the following databases.
The CATH release 1.7 which contains 15802 protein chains, among which
10906 are classified as single domain. This latter set is called PCs. 
We also used the CATH release 2.0
which contains 20780 protein chains, among which 14389 are single domain (PC2s). 
The SCOP release 1.53 which contains 20021 protein chains,
among which 15375 are single domain (PSs).
The FSSP release from Jan 14, 2001 which contains 22660 protein chains (PF).
The FSSP proteins are grouped into 2494 homology classes such that
within a class the sequence similarity is above 25\%.
One protein per class is selected as representative and
we call PFr the set of all representatives.
All the protein sets and their sizes are listed in Table \ref{tab:prot_sets}.

\subsection*{The Classification by Optimization (CO) method}

The classification scheme that we used is based on the minimization
of a particular cost function, defined as follows (for the case of the
prediction of CATH topology; a similar definition holds for SCOP folds);
Each protein is assigned an integer number $c_i$, 
describing its topology (1 to 305).
We assign to proteins with known classification the value of $c(i)$ 
determined by their CATH classification.
To the yet unclassified proteins we assign initially 
random values from 1 to 305.
A cost is calculated for each configuration 
${\cal C}=\SetOf{c_i}$ of topologies
which penalizes the assignment of different topologies
to any pair of proteins.
The value of this penalty is chosen to be the similarity measure 
$Z_{ij}$ between proteins $i$ and $j$;
the higher the similarity $Z_{ij}$, the more costly it is to place proteins
$i$ and $j$ in different topologies.
The cost function is defined as the sum of
penalties for all protein pairs $\ij$,
\be
\label{eq:cost}
E({\cal C})= \sum_{\ij} Z_{ij} \left[ 1-\delta(c_i,c_j) \right] \;.
\ee
The classification problem is stated as finding the minimal cost 
configuration of the unclassified proteins, while keeping 
the topologies (\ie the $c_i$ values) of the classified proteins fixed.
This problem corresponds to finding the ground state of a 
random field Potts ferromagnet.

We search for a classification $C$ of minimal cost
by an iterative greedy algorithm described in detail elsewhere \cite{getz98}
The algorithm identifies at which iteration, 
if any, it performed a heuristic decision.
For low fractions of unknown topologies
the algorithm usually reaches the global minimum of the cost function.

\subsection*{Bounds on the success rate of the prediction}
In this section we establish a statistical 
upper bound for the prediction success rate
relevant to a family of prediction algorithms.

The Z-matrix can be reinterpreted as a {\em weighted graph};
each vertex in the graph represents a protein and the weights on the edges
connecting two vertices are the corresponding Z-scores.
Edges with $Z<2.0$ are absent from the graph.
Following this representation, we define two proteins
as {\em neighbors} if they are connected by an edge.
By analyzing the connectivity properties of set PC
we make inferences about our predictive power.

One can characterize the FSSP-based neighborhood of a protein
according to the CATH classification
of itself and its neighbors. Every protein must belong to one of four
categories: \\
(A) ``Island'' - The protein has no neighbors; \\
(B) ``Colony'' - It has no neighbors of its own kind; \\
(C) ``Border'' - It has neighbors of its own kind as well
as of other kinds; \\
(D) ``Interior'' - The protein has only neighbors of its own kind. \\
Using these definitions we can arrange the proteins of PC
in groups according to their neighborhood category
at the class, architecture and topology levels.
The distribution of the proteins among these groups can be used to calculate
an upper bound for the CO method, if we assume that the set of 
unclassified proteins has the same distribution as the classified ones. 
For example, islands cannot be classified and are therefore rejected.
Colonies are bound to be misclassified since none of their neighbors give a clue
on their type. Since the fraction of proteins in each category was estimated on
the basis of a sample, it can be interpreted only as a {\it statistical} upper
bound. 

We consider the set PFCs to obtain a first type of upper bound
for the success rate of the CO method.
This set (see Table \ref{tab:prot_sets}) 
is formed by 10541 proteins, among which
5\% are islands, a negligible fraction (0.2\%) are colonies, 
6\% are borders and 88\% are interiors.
Therefore the upper bound that we found is about 95\% 
for predicting the topology level in CATH.

The actual prediction performed in this work is done on the set \pfc~
which is formed by the 7431 proteins that are in FSSP (14 Jan 2001)
but not in CATH1.7 (see Table \ref{tab:prot_sets}).
Within \pfc~ there is a subset of 1617 (about 22\%) 
proteins which are either islands or super-islands, i. e.
they are connected only with other proteins in the subset and therefore
they have no connection to proteins with known classification.
Thus, the upper bound for this second type of prediction is about 78\%.

\subsection*{Evaluating a classification prediction algorithm}

Since an algorithm can output either
a predicted classification or a ``rejection'',
if it does not have any prediction,
one has to estimate two probabilities:
$P_{success}$ and $P_{reject}$.
Robust estimation of these parameters is produced by
cross-validation a procedure which consists in
averaging over many ($T$) randomly sampled test trials.
In each trial, the set is divided into two subsets;
one is used for {\em training} the algorithm and the other set,
of $N_{test}$ proteins, is used to {\em test}
the algorithm by comparing its prediction to the true classification.
The probability estimates are given by
\bea
& \hat{P}_{success} & = 1/T \sum_{t=1}^{T} \frac{N_{success}}{N_{test}} \\
& \hat{P}_{non-reject} & = 1-\hat{P}_{reject}= 1/T \sum_{t=1}^{T} \frac{N_{test}-N_{reject}}{N_{test}} \\
\eea
Another figure of merit, the purity $P_{pure}$,
is the probability of correctly classifying
non-rejected proteins. It is estimated by
\be
\hat{P}_{pure}=\frac{\hat{P}_{success}}{1-\hat{P}_{reject}} \;\;.
\ee



\begin{thebibliography}{10}

\bibitem{holm96}
Holm L and Sander C.
\newblock {Mapping the protein universe}.
\newblock Science 1996;  273:595--602.

\bibitem{thornton99}
Thornton JM, Orengo CA, Todd AE, and Pearl FMG.
\newblock {Protein folds, functions and evolution }.
\newblock J. Mol. Biol. 1999;  293:333--342.

\bibitem{sali98}
\v{S}ali A.
\newblock {100,000 protein structures for the biologist}.
\newblock Nature Struct. Biol. 1998;  5:1029--1032.

\bibitem{marti00}
Mart\'{\i}-Renom MA, Ashley AC, Fiser A, Sanchez R, Melo F, and \v{S}ali A.
\newblock {Comparative protein structure modeling og genes and genomes }.
\newblock Annu. Rev. Biophys. Biomol. Struct. 2000;  29:291--325.

\bibitem{heger00}
Heger A and Holm L.
\newblock {Towards a covering set of protein familiy profiles }.
\newblock Prog. Biophys. Mol. Biol. 2000;  73:321--337.

\bibitem{bowie91}
Bowie JU, L\"uthy R, and Eisenberg D.
\newblock {A method to identify protein sequences that fold into a known
  three-dimensional structure}.
\newblock Science 1991;  253:164--170.

\bibitem{jones92}
Jones DT, Taylor WR, and Thornton JM.
\newblock {A new approach to protein fold recognition }.
\newblock Nature 1992;  358:86--89.

\bibitem{fisher96}
Fisher D, Rice D, Bowie JU, and Eisenberg D.
\newblock Assigning amino acid sequences to 3-dimensional protein folds.
\newblock FASEB J. 1996;  10:126--136.

\bibitem{gerstein97}
Gerstein M and Levitt M.
\newblock A structural census of the current population of protein sequences.
\newblock Proc. Natl. Acad. USA 1997;  94:11911--11916.

\bibitem{murzin96}
Murzin AG.
\newblock Structural classification of proteins: new superfamilies.
\newblock Curr. Opin. Struct. Biol. 1996;  6:386--394.

\bibitem{blundell00}
Blundell TL and Mizuguchi K.
\newblock {Structural genomics: an overview}.
\newblock Prog. Biophys. Mol. Biol. 2000;  73:289--295.

\bibitem{finkelstein97}
Finkelstein AV.
\newblock {Can protein unfolding simulate protein folding?}
\newblock Prot. Eng. 1997;  10:843--845.

\bibitem{simons99}
Simons KT, Bonneau R, Ruczinski I, and Baker D.
\newblock {Ab initio protein structure prediction of CASP III targets using
  ROSETTA }.
\newblock Proteins 1999;  37:S3 171--176.

\bibitem{bernstein77}
Bernstein F, Koetzle T, Williams G, Meyer EJ, Brice M, Rodgers J, Kennard O,
  Shimanouchi T, and Tasumi M.
\newblock {The protein data bank: A computer-based archival file for
  macromolecular structures}.
\newblock J. Mol. Biol. 1977;  112:535--542.

\bibitem{holm97}
Holm L and Sander C.
\newblock {Dali/FSSP classification of three-dimensional protein folds }.
\newblock Nucleic Acids Res. 1997;  25:231--234.

\bibitem{orengo97}
Orengo CA, Michie AD, Jones S, Jones DT, Swindells MB, and Thornton JM.
\newblock {CATH - a hierarchic classification of protein domain structures }.
\newblock Structure 1997;  5:1093--1108.

\bibitem{loconte00}
Conte LL, Ailey B, Hubbard TJP, Brenner SE, Murzin AG, and Chothia C.
\newblock {SCOP: a structural classification of proteins database }.
\newblock Nucleic Acids Res. 2000;  28:257--259.

\bibitem{mizuguchi98}
Mizuguchi K, Deane CM, Blundell TL, and Overington JP.
\newblock {HOMSTRAD: a database for protein structure alignments for homologous
  families }.
\newblock Prot. Sci. 1998;  7:2469--2471.

\bibitem{gibrat96}
Gibrat JF, Madej T, and Bryant SH.
\newblock Surprising similarities in structure comparison.
\newblock Curr. Opin. Struct. Biol. 1996;  6:377--385.

\bibitem{siddiqui95}
Siddiqui AS and Barton GJ.
\newblock {Continuous and discontinuous domains: an algorithm for the automatic
  generation of reliable protein domain definitions }.
\newblock Prot. Sci. 1995;  4:872--884.

\bibitem{getz98}
Getz G.
\newblock {Clustering and Classification of Protein Structures}.
\newblock M. Sc. Thesis, Tel-Aviv University,  1998; .

\bibitem{hadley99}
Hadley C and Jones DT.
\newblock {A systematic comparison of protein structure classifications: SCOP,
  CATH and FSSP}.
\newblock Structure 1999;  7:1099--1112.

\bibitem{holm94}
Holm L and Sander C.
\newblock {The FSSP database of structurally aligned protein fold families}.
\newblock Nucleic Acids Res. 1994;  22:3600--3609.

\bibitem{dietmann01}
Dietmann S, Park J, Notredame C, Heger A, Lappe M, and Holm L.
\newblock {A fully automatic evolutionary classification of protein folds: Dali
  Domain Dictionary version 3 }.
\newblock Nucleic Acids Res. 2001;  29:55--57.

\bibitem{jain88}
Jain AK and Dubes RC.
\newblock Algorithms for Clustering Data.
\newblock Prentice--Hall, Englewood Cliffs,  1988; .

\bibitem{levitt76}
Levitt M and Chothia C.
\newblock {Structural patterns in globular proteins}.
\newblock Nature 1976;  261:552--558.

\bibitem{taylor89}
Taylor WR and Orengo CA.
\newblock {Protein Structure Alignment }.
\newblock J. Mol. Biol. 1989;  208:1--22.

\bibitem{orengo92}
Orengo CA, Brown NP, and Taylor WR.
\newblock {Fast Structure Alignment for Protein Databank Searching }.
\newblock Proteins 1992;  14:139--167.

\bibitem{bray00}
Bray JE, Todd AE, Pearl FMG, Thornton JM, and Orengo CA.
\newblock {The CATH Dictionary of Homologous Superfamilies: a consensus
  approach to analyze distant structural homologues}.
\newblock Prot. Eng. 2000;  13:153--165.

\bibitem{brenner00}
Brenner SE, Koehl P, and Levitt M.
\newblock {The ASTRAL compendium for protein structure and sequence analysis}.
\newblock Nucleic Acids Res. 2000;  28:254--256.

\bibitem{getz01web}
{http://www.weizmann.ac.il/physics/complex/compphys/f2cs/index.html}.

\bibitem{swindells98}
Swindells MB, Orengo CA, Jones DT, Hutchinson EG, and Thornton JM.
\newblock {Contemporary approaches to protein structure classification}.
\newblock BioEssays 1998;  20:884--891.

\bibitem{islam95}
Islam SA, Luo J, and Sternberg MJE.
\newblock {Identification and analysis of domains in proteins}.
\newblock Prot. Eng. 1995;  8:513--525.

\bibitem{swindells95}
Swindells MB.
\newblock {A procedure for detecting structural domains in proteins}.
\newblock Prot. Sci. 1995;  4:103--112.

\bibitem{sowdahamini96}
Sowdahamini R, Rufino SD, and Blundell TL.
\newblock {Nuclear dynamics and electronic transition in a photosynthetic
  reaction center}.
\newblock J. Am. Chem. Soc. 1997;  119:3948--3958.

\end{thebibliography}


\newpage
\vspace*{-2.0cm}
\begin{table}
\begin{center}
\leavevmode
\begin{tabular}{ |l|l|}
\hline
Abbreviation & Definition \\
\hline
3Dee & Database of Protein Domain Definitions \\
ASTRAL &The ASTRAL Compendium for Sequence and Structure Analysis \\
CATH & Protein Structure Classification \\
CO & Classification by Optimization \\
DALI & Protein Structure Comparison by Alignment of Distance Matrices \\
DHS & Dictionary of Homologous Superfamilies\\
FSSP & Fold classification based on Structure-Structure alignment of Proteins \\
HOMSTRAD & HOMologous STRucture Alignment Database\\
MMDB & Molecular Modeling DataBase\\
PDB & Protein Data Bank\\
SCOP & Structural Classification of Proteins\\
SSAP & Structure Comparison Algorithm\\
\hline
\end{tabular}
\vspace{0.2cm}
\caption{Abbreviations and definitions.}
\label{tab:abbrev}
\end{center}
\end{table}
\vspace{10truemm}

\vspace*{-2.0cm}
\begin{table}
\begin{center}
\leavevmode
{\tiny
\begin{tabular}{|l|l|l|}
\hline
Original Classification & Assigned classification & percent of cases \\
\hline
\hline
 {\bf Mainly Alpha }        &                             &        \\
\hline
 1.10 Orthogonal Bundle     &  1.10 Orthogonal Bundle     & 96.3\% \\ 
                            &  Reject                     &  3.3\% \\ 
\hline
 1.20 Up-down Bundle        &  1.20 Up-down Bundle        & 97.7\% \\ 
                            &  4.10 Irregular             &  1.2\% \\ 
                            &  1.10 Orthogonal Bundle     &  0.7\% \\ 
\hline
 1.25 Horseshoe             &  1.25 Horseshoe             & 100.0\% \\ 
\hline
 1.50 Alpha/alpha barrel    &  1.50 Alpha/alpha barrel    & 100.0\% \\ 
\hline
\hline
 {\bf Mainly Beta}          &                             &        \\
\hline
 2.10 Ribbon                &  2.10 Ribbon                & 93.9\% \\ 
                            &  Reject                     &  5.7\% \\ 
\hline
 2.20 Single Sheet          &  2.20 Single Sheet          & 97.2\% \\ 
                            &  Reject                     &  2.3\% \\ 
\hline
 2.30 Roll                  &  2.30 Roll                  & 97.2\% \\ 
                            &  Reject                     &  2.1\% \\ 
                            &  3.10 Roll                  &  0.7\% \\ 
\hline
 2.40 Barrel                &  2.40 Barrel                & 91.0\% \\ 
                            &  Reject                     &  8.8\% \\ 
\hline
 2.50 Clam                  &  2.50 Clam                  & 94.4\% \\ 
                            &  2.40 Barrel                &  5.6\% \\ 
\hline
 2.60 Sandwich              &  2.60 Sandwich              & 86.1\% \\ 
                            &  Reject                     & 13.9\% \\ 
\hline
 2.70 Distorted Sandwich    &  2.70 Distorted Sandwich    & 96.1\% \\ 
                            &  2.60 Sandwich              &  3.9\% \\ 
\hline
 2.80 Trefoil               &  2.80 Trefoil               & 100.0\% \\ 
\hline
 2.90 Orthogonal Prism      &  2.90 Orthogonal Prism      & 100.0\% \\ 
\hline
 2.100 Aligned Prism        &  2.100 Aligned Prism        & 100.0\% \\ 
\hline
 2.102 3-layer Sandwich     &  2.102 3-layer Sandwich     & 78.6\% \\ 
                            &  2.30 Roll                  & 21.4\% \\  
\hline
 2.110 4 Propellor          &  2.110 4 Propellor          & 100.0\% \\ 
\hline
 2.120 6 Propellor          &  2.120 6 Propellor          & 96.1\% \\ 
                            &  Reject                     &  3.9\% \\ 
\hline
 2.130 7 Propellor          &  2.130 7 Propellor          & 100.0\% \\ 
\hline
 2.140 8 Propellor          &  2.140 8 Propellor          & 85.3\% \\ 
                            &  Reject                     & 14.7\% \\ 
\hline
 2.160 3 Solenoid           &  2.160 3 Solenoid           & 100.0\% \\ 
\hline
 2.170 Complex              &  2.170 Complex              & 83.3\% \\ 
                            &  2.60 Sandwich              &  8.6\% \\ 
                            &  Reject                     &  8.0\% \\ 
\hline
\hline
 {\bf Mixed Alpha-Beta}     &                             &        \\
\hline
 3.10 Roll                  &  3.10 Roll                  & 99.9\% \\ 
\hline
 3.20 Barrel                &  3.20 Barrel                & 100.0\% \\ 
\hline
 3.30 2-Layer Sandwich      &  3.30 2-Layer Sandwich      & 93.5\% \\ 
                            &  Reject                     &  6.0\% \\ 
\hline
 3.40 3-Layer(aba) Sandwich &  3.40 3-Layer(aba) Sandwich & 96.1\% \\ 
                            &  Reject                     &  3.8\% \\ 
\hline
 3.50 3-Layer(bba) Sandwich &  3.50 3-Layer(bba) Sandwich & 72.1\% \\ 
                            &  Reject                     & 27.2\% \\ 
                            &  3.30 2-Layer Sandwich      &  0.7\% \\ 
\hline
 3.60 4-Layer Sandwich      &  3.60 4-Layer Sandwich      & 99.7\% \\ 
\hline
 3.70 Box                   &  3.70 Box                   & 100.0\% \\ 
\hline
 3.75 5-stranded Propeller  &  3.75 5-stranded Propeller  & 100.0\% \\  
\hline
 3.80 Horseshoe             &  3.80 Horseshoe             & 100.0\% \\ 
\hline
 3.90 Complex               &  3.90 Complex               & 97.9\% \\ 
                            &  Reject                     &  0.7\% \\ 
\hline
\hline
 {\bf Few Secondary Structures} &                             &        \\
\hline
 4.10 Irregular             &  4.10 Irregular             & 90.8\% \\ 
                            &  Reject                     &  8.3\% \\ 
                            &  1.20 Up-down Bundle        &  0.8\% \\ 
\hline
\end{tabular}
}
\vspace{0.2cm}
\caption{Summary of a ``confusion table''.
This table summarizes the results about the assignments of the
CATH architecture for proteins that have already a CATH classification.
Only cases which occur more than $0.5\%$ are listed.
These figures were calculated using 100 cross-validation runs 
at 20\% dilution. 
}
\label{tab:confusion}
\end{center}
\end{table}
\vspace{10truemm}

\newpage
\begin{table}
\begin{center}
\leavevmode
\begin{tabular}{ l|l|l }
\hline
Name & Description & Size of set\\
\hline
PF & all chains in FSSP (14 Jan, 2001) & 22660 \\
PFr & representative chains in FSSP (14 Jan, 2001) & 2494 \\
PC & chains in CATH v1.7 & 15802\\ 
PCs & single domain chains in CATH v1.7 & 10906 \\ 
PC2 & chains in CATH v2.0 & 20780 \\ 
PC2s & single domain chains in CATH v2.0 & 14389 \\ 
PS & chains in SCOP 1.53 & 20021 \\ 
PSs & single domain chains in SCOP 1.53 & 15375 \\ 
PCsSs & single domains chains in SCOP 1.53 and CATH v1.7 & 10197 \\
PFrCs & single domain chains in CATH that are representatives FSSP ($PFr \cap PCs$)& 860 \\
PFrSs & single domain chains in SCOP that are representatives FSSP ($PFr \cap PSs$)& 1626 \\
PFCs & chains in FSSP and single domain in CATH v1.7 & 10541 \\
PFSs & chains in FSSP and single domain in SCOP 1.53 & 14716 \\
\pfc & chains in FSSP and not in CATH v1.7 & 7431 \\
\pfs & chains in FSSP and not in SCOP 1.53 & 3451 \\
\hline
\end{tabular}
\vspace{0.2cm}
\caption{Protein sets and their sizes.}
\label{tab:prot_sets}
\end{center}
\end{table}
\vspace{10truemm}

\newpage
\begin{table}
\begin{center}
\leavevmode
\begin{tabular}{ |l|l|l|l|l|l|l|l|l|l|l|l|l|l|l|l|l|}
\hline
Chain id & \multicolumn{4}{|c|}{CATH v1.7} & \multicolumn{4}{|c|}{CATH v2.0} & \multicolumn{3}{|c|}{{\bf CATH Prediction}} & \multicolumn{3}{|c|}{SCOP 1.53} & \multicolumn{2}{|c|}{{\bf SCOP Prediction}} \\ \cline{2-17}
 & \# & C & A & T & \# & C & A & T & {\bf C} &{\bf A} & {\bf T} & \# & C & F & {\bf C} & {\bf F} \\ \hline
1cuoa & -1 & & & & 1 & 2 & 60 & 40 & {\bf 2} & {\bf 60} & {\bf 40} & -1 &  & & {\bf 2} & {\bf 5} \\ \hline
\end{tabular}
\vspace{0.2cm}
\caption{The search result when submitting ``1cuoA'' to the web site 
http://www.weizmann.ac.il/physics/complex/compphys/f2cs/.
This protein was classified by neither CATH v1.7 nor SCOP 1.53 
which are the basis of our predictions. We predicted it to belong 
to CATH topology 2.60.40 and SCOP fold 2.5. Later it was indeed 
classified by CATH v2.0 as 2.60.40. The -1 in both CATH v1.7 and 
SCOP 1.53 represents that it was not classified by them.
}
\label{tab:1cuoA}
\end{center}
\end{table}
\vspace{10truemm}

\newpage
\vspace{10truemm}

\newpage
\vspace{10truemm}

\newpage
\vspace{10truemm}

\newpage
\vspace{10truemm}

\newpage

\end{document}